\pdfoutput=1
\documentclass[twocolumn,english,aps,superscriptaddress,prb]{revtex4-1}

\usepackage{babel}
\usepackage{amsmath}
\usepackage{amssymb}
\usepackage{graphicx}

\begin{document}

\title{From Kosterlitz-Thouless to Pokrovsky-Talapov transitions in spinless fermions and spin chains with next-nearest neighbour interactions}

\author{Natalia Chepiga}
\affiliation{Kavli Institute of Nanoscience, Delft University of Technology, Lorentzweg 1, 2628 CJ Delft, The Netherlands}

\date{\today}
\begin{abstract} 
We investigate the nature of the quantum phase transition out of charge-density-wave phase in the spinless fermion model with nearest- and next-nearest-neighbor interaction at one-third filling. Using an extensive Density Matrix Renormalization Group (DMRG) simulations we show that the transition changes it nature. We show that for weak next-nearest-neighbor coupling the transition is of  Kosterlitz-Thouless type in agreement with bosonisation predictions. We also provide strong numerical evidences that for large next-nearest-neighbor repulsion the transition belongs to the Pokrovsky-Talapov univerality class describing a non-conformal commensurate-incommensurate transition. Finally, we argue that the change of the nature of the transition is a result of incommensurability induced by frustration and realized even at zero doping. The implications in the context of XXZ chain with next-nearest-neighbor Ising interaction is briefly discussed.
\end{abstract}

\maketitle


\section{Introduction}

Understanding the nature of quantum phase transitions in low-dimensional systems is one of the central topics in condensed matter physics\cite{giamarchi,tsvelik,sachdev_QPT}. The conjecture of universality classes allows to investigate phase transitions on a simple lattice models. A paradigmatic example that have appeared in many contexts over the decades is a model of interacting spinless fermions in 1D\cite{giamarchi}. Their applications range from solving spin models through Jordan-Wigner transformation\cite{pfeuty} to studying the commensurate melting of classical 2D and later quantum 1D models\cite{HuseFisher1984,Den_Nijs,GIAMARCHI1997975,fendley,kibble_zureck,prl_chepiga,dalmonte}. 
The models with competing nearest- (NN) and next-nearest-neighbor (NNN) interactions has been studied intensely over the years\cite{PhysRevB.56.R1645,PhysRevB.56.12939,PhysRevB.64.033102,Duan_2011,roux}. Initially formulated as a toy model for the long-range Coulomb interaction, it has been soon realized that the model despite its simplicity has a very rich phase diagram hosting in particular the Luttinger liquid phase, the charge density waves at half- and one-third fillings, and paired phases\cite{PhysRevLett.111.165302,PhysRevB.92.045106,roux}. The full phase diagram however is far from being complete and there are number questions that remain open. For instance, the properties of the highly entropic phase reported recently\cite{roux} or the nature of quantum phase transitions that will be the main focus of this paper. 
 Over the last decades, the combination of conformal field theory in 1+1D\cite{cardy,difrancesco} and advanced numerical techniques such as the density matrix renormalization group algorithm (DMRG)\cite{dmrg1,dmrg2,dmrg3,dmrg4} has proven to be extremely powerful in coming up with theoretical predictions for numerous fascinating critical phenomenas.

In this paper we investigate the nature of the quantum phase transition between the charge density wave at one-third filling and the Luttinger liquid phase for a chain of spinless fermions with nearest- and next-nearest-neighbor repulsion. 
The microscopic model is defined by the following Hamiltonian:
\begin{equation}
    H_\mathrm{ferm}=\sum_i-t(c^\dagger_ic_{i+1}+\mathrm{h.c.})+U_1 n_i n_{i+1}+U_2 n_i n_{i+2},
    \label{eq:ferm}
\end{equation}
where $c_i^\dagger,c_i$ are fermionic creation and annihilation operators, $t$ is a hoping amplitude and $U_{1,2}$ are nearest- and next-nearest-neighbor coupling constants correspondingly. The model can be reformulated in terms of hard-core bosons and the Hamiltonian takes essentially the same form. By means of Jordan-Wigner transformation the model can also be rewritten in terms of spin-1/2 operators with the following Hamiltonian: 
\begin{equation}
    H_\mathrm{spin}=\sum_i-t(S^+_iS^-_{i+1}+S^-_iS^+_{i+1})+U_1 S^z_i S^z_{i+1}+U_2 S^z_i S^z_{i+2},
    \label{eq:spin}
\end{equation}
 where $S^\pm_i=S^x_i\pm iS^y_i$. The Hamiltonian (\ref{eq:ferm}) conserves the total number of fermions, while the spin version of the Hamiltonian (\ref{eq:spin}) conserves the total magnetization. It is therefore natural to study the phase diagram of these models at a fixed filling or fixed magnetization. In the non-interacting case $U_1=U_2=0$ the ground-state can be described by the Luttinger liquid with the Luttinger liquid parameter $K=1$. Repulsive interactions does not destroy the Luttinger liquid phase immediately but the Luttinger liquid parameter takes values $K<1$.  Recently it has been shown that at one-third filling a large portion of the phase diagram is occupied by the density wave phase that spontaneously breaks translation symmetry with every third site occupied by a fermion\cite{roux}. 
 In the spin language this corresponds to the total magnetization  $S^z_\mathrm{tot}=-(N-1)/6$ and the ground state of the form $\uparrow \downarrow \downarrow$. According to the theory of Mott transitions that takes place as a function of coupling while the filling is fixed (often referred to as Mott-U transitions\cite{GIAMARCHI1997975}) one can expect the transition between the charge-density-wave phase and the Luttinger liquid phase to be of the Kosterlitz-Thouless type\cite{Kosterlitz_Thouless}. At the transition the Luttinger liquid exponent reaches its critical value $K^c=2/9$ obtained with bosonisation\cite{roux}. But how robust is this prediction against multiple competing interactions?
 
Recently, it has been shown that the Luttinger liquid phase can be stabilized for a model with strong nearest and next-nearest-neighbor repulsion up to the Luttinger liquid parameter $K^c=1/9$. The result is exact in the limit of next-nearest-neighbor blockade\cite{1999cond.mat..4042A}, but it is expected to hold even for large enough but finite interaction strength\cite{verresen}. In this case the transition out of charge-density wave phase belongs to the Pokrovsky-Talapov universality class\cite{Pokrovsky_Talapov}.  The stability of the Luttinger liquid has been studied as a function of chemical potential $\mu$, that in Giamarchi's notations\cite{GIAMARCHI1997975} corresponds to the Mott-$\delta$ transition, i.e. the one that takes place at fixed coupling while the filling is tuned by $\mu$. However, since the critical Luttinger liquid parameter is smaller than the one at the Kosterlitz-Thouless transition predicted for small $U_2$ one might expect an extension of the Luttinger liquid phase beyond $K^c=2/9$ for strong next-nearest-neighbor repulsion. 

\begin{figure}[t!]
\centering 
\includegraphics[width=0.4\textwidth]{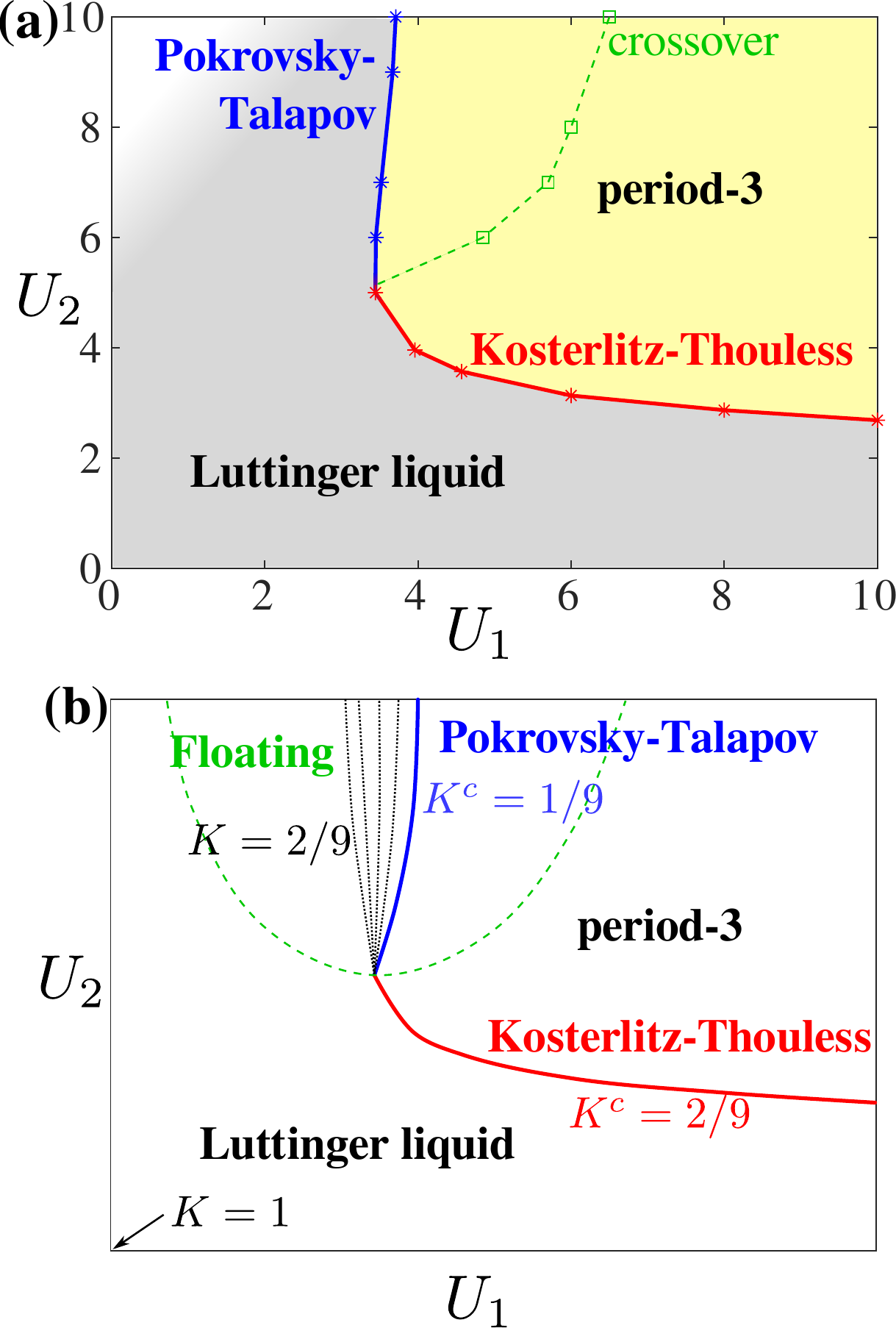}
\caption{Phase diagram as a function of nearest and next nearest neighbor repulsion for the one-third filling. (a) Phase diagram obtained with DMRG simulations. For $U\lesssim5$ the transition is in the Kosterlitz-Thouless universality class; for $U\gtrsim 5$ the transition is of Pokrovsky-Talapov type. Green squares mark the location of the kinks in the correlation length, dashed line is a guide to eyes. White region corresponds to the possible highly entropic phase\cite{roux} which is out of the scope of the present study. (b) Sketch of the main features of the Luttinger liquid phase on the phase diagram presented in (a). The Luttinger liquid parameter is $K=1$ in the non-interacting case $U_1=U_2=0$ and decreases upon approaching the boundary of the period-three phase. At the Kosterlitz-Thouless transition the Luttinger liquid parameter takes the critical value $K^c=2/9$. At the Pokrovsky-Talapov transition the corresponding  critical value is $K^c=1/9$. Equal-$K$ lines with $1/9<K<2/9$ are expected to collapse at the point where the transition changes its nature. The crossover line (green dashed) is expected to continue in the critical phase separating the commensurate Luttinger liquid form the incommensurate floating phase.}
\label{fig:phasediag}
\end{figure}

In this paper we show that the nature of the quantum transition out of the period-three phase changes from the Kosterlitz-Thouless type realized for weak coupling $U_2$ to the Pokrovsky-Talapov universality class that appears for strong $U_2$ as shown in Fig.\ref{fig:phasediag}. We argue that incommensurability that leads to the commensurate-incommensurate  Pokrovsky-Talapov transition is a result of a frustration induced by the repulsion and a filling constraint. This results in the appearance of the floating phase - a region of the Luttinger liquid phase with the local density in the finite-size system oscillating around its fixed value $1/3$ with the wave-vector noticeably different form the commensurate value $q=2\pi/3$.  The rest of the paper is organized as follows. In section \ref{sec:method} we will provide the technical details of the used numerical method. In section \ref{sec:KT} we numerically verify the prediction for Kosterlitz-Thouless transition for small $U_2$. In the next section  \ref{sec:PT} we provide numerical evidence for the Pokrovsky-Talapov transition and demonstrate the emergence of the incommensurate floating phase. Finally, in the section \ref{sec:conclusion} we summarize the results and put them into a perspective.

\section{Numerical method}
\label{sec:method}
 
 Numerical simulations have been performed with the density matrix renormalization group (DMRG) algorithm\cite{dmrg1,dmrg2,dmrg3,dmrg4} for the spin model of Eq.\ref{eq:spin}.  Without loss of generality the hopping amplitude is set to $t=1$ throughout a paper.  The results were obtained for chains with up to $N=3601$ sites with open boundary conditions, keeping  up to $D=3000$ states and discarding singular values below $10^{-8}$. This allow to converge the ground-state energy with the error well below $10^{-7}$. Thus high accuracy turns out to be a prerequisite to observe the floating phase on a finite size chains with $N=301$ and $601$ sites (see for instance Fig.\ref{fig:Float}).  
  In order to realize 1/3 filling of the fermionic model, the algorithm is constrained to the sector with total magnetization $S^z_\mathrm{tot}=-(N-1)/6+1/2$ with $N=3k+1$,  $k\in\mathbb{Z}$. The boundary conditions are fixed by polarizing the edge spins in $z$-direction.
  
\begin{figure}[t!]
\centering 
\includegraphics[width=0.5\textwidth]{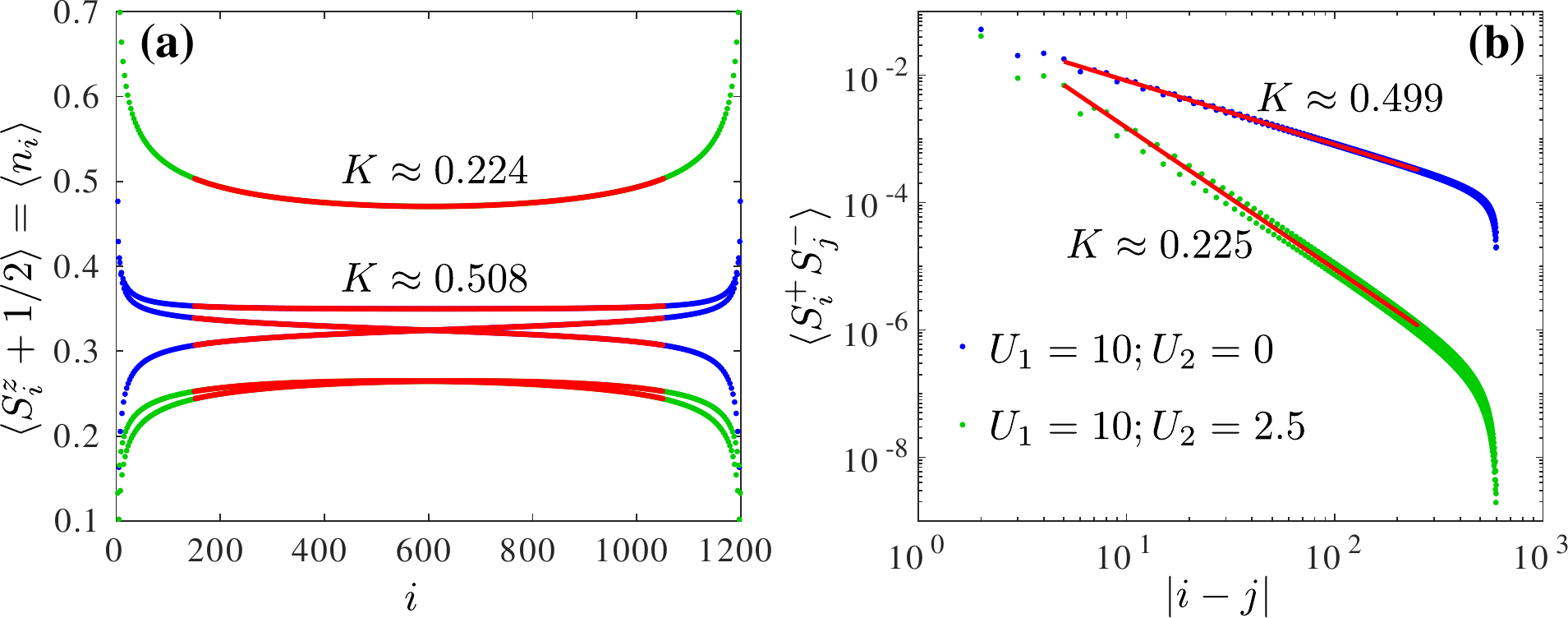}
\caption{Extraction of the Luttinger liquid exponent $K$ by fitting (a) the Friedel oscillations of the local density on a finite chain and (b) the transverse component of the spin-spin correlations. The results are shown for $U_1=10$ at $U_2=0$ (blue) and at $U_2=2.5$ (green) (in the thermodynamic limit the critical point is located at $U_2\approx2.462$). The results of the fit are shown in red. }
\label{fig:fitexample}
\end{figure}

  Most of the results were obtained with quantities that do not depend on the statistics. The correlation length $\xi$ is extracted inside the gapped period-three phase by fitting an exponential decay of the correlation function $\langle S^z_i S^z_j\rangle$. Up to a constant this is equivalent to $\langle n_in_j\rangle$ of the fermion model. The Luttinger liquid parameter $K$ is extracted by fitting the profile of the local magnetization $\langle S^z_i\rangle$. This is equivalent to the local density profile $\langle n_i\rangle$. Fixed boundary conditions act as an impurity and lead to Friedel oscillations. According to the boundary conformal field theory the profile takes the following form:
  $$S^z_j\propto \frac{\cos(q j)}{[(N/\pi) \sin (\pi j/N)]^{K}}.$$
One can benchmark the value of the  Luttinger liquid exponent $K$ obtained with Friedel oscillations by comparing it to the slope of the correlation function of the transverse components of spins $\langle S^+_iS^-_j\rangle$. This is the only statistics sensitive quantity used in this paper. This correlation function is much easier to compute and to fit in the spin language than in the fermionic one. In Fig.\ref{fig:fitexample} I provide some examples of the fits. In general the two methods give similar results and from now on we only present the results obtained with Friedel oscillations. The error bars are estimated by fitting different intervals of the data points discarding from 5\% to 25\% of spins at each edge. Finally, the incommensurate wave-vector $q$ has been obtained by fitting the Friedel oscillations inside the floating phase (some examples are presented in Fig.\ref{fig:Float}).

\section{Kosterlitz-Thouless transition}
\label{sec:KT}
 
\begin{widetext}

   \begin{figure}[h!]
\centering 
\includegraphics[width=0.99\textwidth]{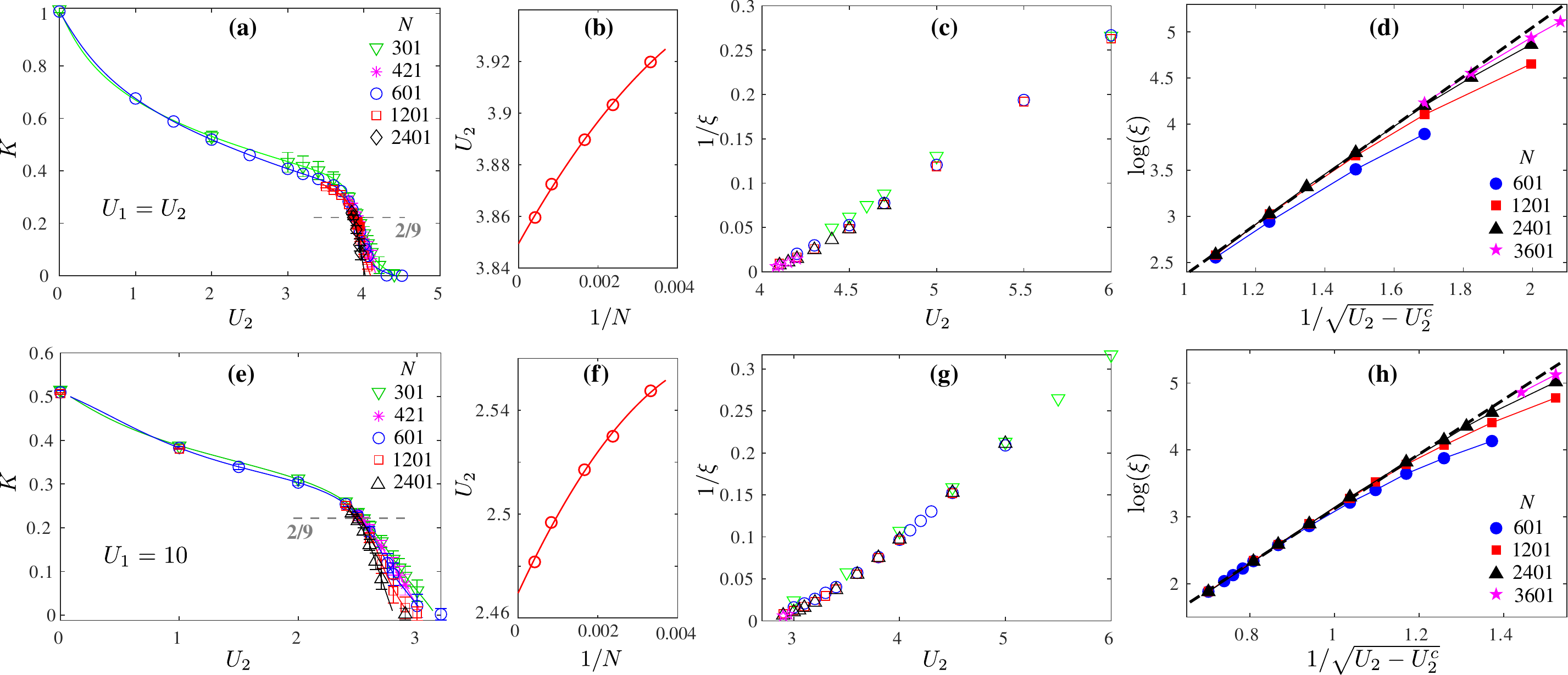}
\caption{Numerical results for the Kosterlitz-Thouless transition along (a-d) $U_1=U_2$ diagonal cut and (e-h) $U_1=10$ vertical cut. (a),(e) Effective values of the Luttinger liquid parameter $K$ extracted by fitting finite-size Friedel oscillation profiles. (b),(f) Finite-size extrapolation of the location of the finite-size critical point associated with $K^c=2/9$ towards the thermodynamic limit. Red circles are data, solid lines are polynomial in $1/N$ fits. Extrapolated critical values are (a) $U_1^c=U_2^c\simeq3.849$ and (f) $U_1=10$, $U_2^c\simeq2.469$. (c),(g) Inverse of the correlation length extracted from the density-density correlations in the period-three phase. (d),(h)  Exponential divergence of the correlation length in the period-three phase as a function of a distance to the transition. Dashed lines are linear fit of (b) five and (c) seven points the farthest from the transition for $N=2401$.   }
\label{fig:U1U2}
\end{figure}

\end{widetext}

Let us first focus on the Kosterlitz-Thouless transition and let us start with the line $U_1=U_2$. The results for the Luttinger liquid exponent $K$ extracted by fitting the Friedel oscillations  are presented in Fig.\ref{fig:U1U2}(a). In the non-interacting case the well known result $K=1$ is recovered. Away from this point the Luttinger liquid exponent decreases in agreement with the repulsive interactions in the system. Close to the transition finite-size effects becomes stronger. Note, that due to exponential divergence of the correlation length typical for the Kosterlitz-Thouless transition, one can extract an effective Luttinger liquid exponent even beyond the transition; this effective exponent is expected to decay to zero in the thermodynamic limit.

 In order to locate the critical point in the thermodynamic limit we first locate the point where the curve $K(U_2)$ for each system size $N$ cross the line $K^c=2/9$ and then extrapolate the obtained values with a polynomial fit in $1/N$ as shown in Fig.\ref{fig:U1U2}(b). The correlation length is computed by fitting exponential decay of the density-density correlations in the period-three phase. At the Kosterlitz-Thouless transition the correlation length is expected to diverge as $\xi\propto \exp{A/\sqrt{U-U^c}}$\cite{Kosterlitz_Thouless}, where $A$ is some non-universal constant.  Fig.\ref{fig:U1U2}(c) show the inverse of the correlation length upon approaching the transition. In Fig.\ref{fig:U1U2}(b) the correlation length is shown as a function of a square-root of a distance to the transition in a semi-log scale. One can see that the scaling systematically approaches the straight line in agreement with the theory prediction. 
 
The results along the cut at $U_1=10$ are organized in a similar way and presented in Fig.\ref{fig:U1U2}(e-h). Like in the previous case one can see a good agreement between numerical data and an expected exponential divergence of the correlation length. These results confirm that for small values of next-nearest-neighbor repulsion $U_2$ the transition belongs to the Kosterlitz-Thouless universality class and the Luttinger liquid exponent takes the value $K^c=2/9$ at the transition.

\section{Pokrovsky-Talapov transition}
\label{sec:PT}

\begin{widetext}
   
   \begin{figure}[t!]
\centering 
\includegraphics[width=0.95\textwidth]{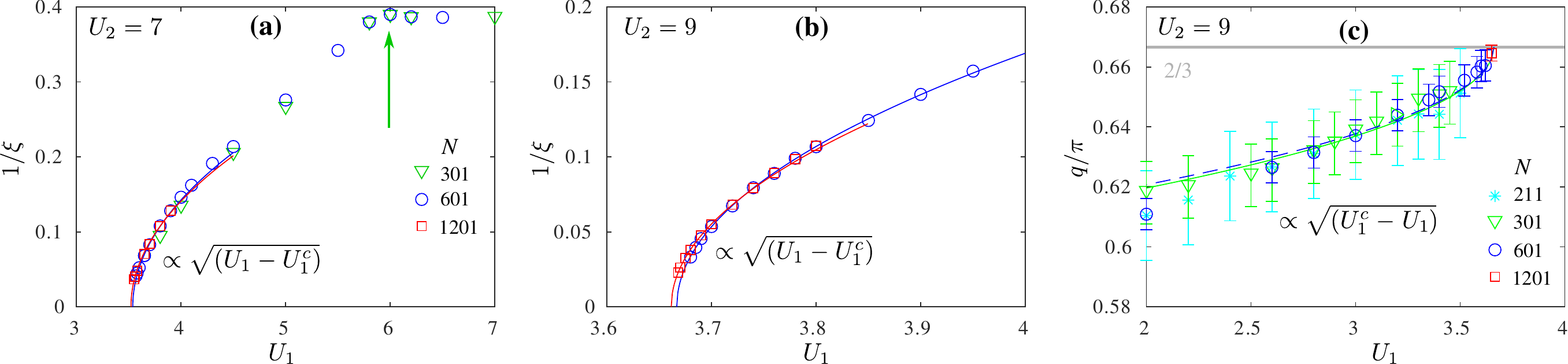}
\caption{ Numerical evidences of the Pokrovsky-Talapov transition. (a),(b) Inverse of the correlation length upon approaching the transition from the period-three phase along (a) $U_2=7$ and (b) $U_2=9$. Symbols are numerical data, lines are fits with $1/\xi\propto (U-U_c)^{\nu}$ with the Pokrovsky-Talapov critical exponent $\nu=1/2$. In (a) one can see the appearance of pronounced kinks indicated with green arrows. The location of the kink corresponds to green squares on the phase diagram of Fig.\ref{fig:phasediag}. (c) Incommensurate wave-vector $q$ as a function of $U_1$ upon approaching the Pokrovsky-Talapov transition. Numerical data (symbols) agree with the theory prediction (lines) $q-2\pi/3 \propto (U_1^c-U_1)^{\bar{\beta}}$ with the Pokrovsky-Talapov critical exponent $\bar{\beta}=1/2$. Green line show the result of the fit for $N=301$ and dashed blue line states for $N=601$. For the fits $U_1^c$ has been fixed to the value obtained in (b) for $N=1201$ sites.}
\label{fig:PT}
\end{figure}

\end{widetext}

Let us now take a look at the boundary of the same period-three phase but for large values of $U_2$. From Fig.\ref{fig:PT}(a),(b) one can see that the inverse of the correlation length upon approaching the transition  vanishes with an infinite slope  in a striking difference to the Kosterlitz-Thouless transition with exponentially diverging correlation length (compare with Fig.\ref{fig:U1U2}(c),(g)). Obtained data points are fit with $1/\xi\propto (U_1-U_1^c)^\nu$ with $\nu=1/2$ the correlation length critical exponent specific for the Pokrovsky-Talapov transition\cite{Pokrovsky_Talapov}. The fits are in spectacular agreement with the numerical data, especially given that there are only two fitting parameters: the location of the critical point $U_1^c$ and the pre-factor.
Slight shift of the critical point is a typical finite-size effect\cite{3boson,2022arXiv220301163M}.

Pokrovsky-Talapov transition is a commensurate-incommensurate transition and the natural question that arises at this stage is how the incommensurability appears in the phase diagram with the fixed commensurate filling?
By looking at the correlation length farther away from the transition one can notice a pronounced kink. Typically, when there is no constraint on the filling, such kinks signal the disorder line separating commensurate and incommensurate regimes. Inside the period-three phase, however, the fixed fillings as well as the long-range order of the gapped phase force the dominant wave-vector to be commensurate $q=2\pi/3$. Within the numerical precision obtained results always agree with this value. Interestingly, if one keeps track of the location of the kink, it turns out that the line crosses the boundary of the period-three phase at $U_2\approx5$ as shown in  Fig.\ref{fig:phasediag}. This agree with the point where the transition changes its nature.

\begin{figure}[t!]
\centering 
\includegraphics[width=0.5\textwidth]{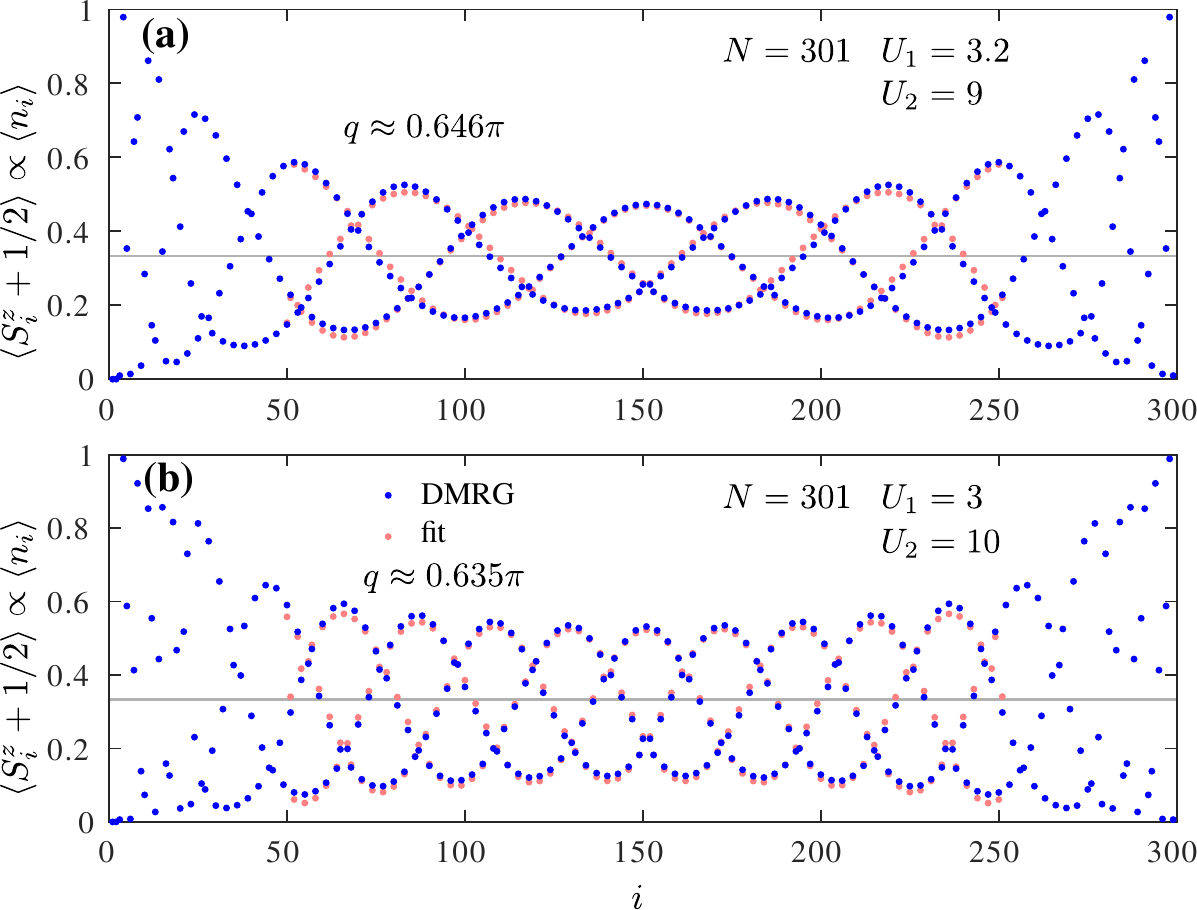}
\caption{ Examples of the local density profiles inside the Luttinger liquid phase for large values of $U_2$. Competition between the one-third filling constraint and next-nearest-neighbor interaction lead to incommensurate oscillations of the local density around the fixed value $1/3$ (gray line). The DMRG data are shown in blue, the results of the fit are shown in pink, an incommensurate wave-vector $q$ obtained from the fit is indicated at each panel. }
\label{fig:Float}
\end{figure}

It is natural to expect the crossover line to continue in the critical phase where it will separates commensurate Luttinger liquid from the floating phase as sketched in Fig.\ref{fig:phasediag}(b). In the critical phase the appearance of the incommensurability can be captured explicitly. Fig.\ref{fig:Float} provides a few examples of the Friedel oscillations profile where the local density fluctuates around $n=1/3$ with the wave-vector noticeably different from $2\pi/3$. For the model written in terms of spin operators in Eq.\ref{eq:spin} it is easy to see that all next-nearest-neighbor bonds cannot be simultaneously minimized for the imposed total magnetization $S^z_\mathrm{tot}=-(N-1)/6$ with commensurate period-3 configuration $\uparrow\downarrow\downarrow$.  The incommensurate fluctuations of local density presented in Fig.\ref{fig:Float} appear as a response to this frustration and give rise to a commensurate-incommensurate nature of the Pokrovsky-Talapov transition at large $U_2$. 

By fitting the Friedel oscillations profile as shown in Fig.\ref{fig:Float} one can get an accurate estimate of the incommensurate wave-vector $q$. The results for $U=9$ are summarized in Fig.\ref{fig:PT}(c). The errorbars are estimated as $\delta q\approx\pi/N$ - the elementary value of the wave-vector that on the entire chain with $N$ sites accumulates into one turn by $\pi$. At the Pokrovsky-Talapov\cite{Pokrovsky_Talapov} transition the wave-vector $q$ is expected to approach its commensurate value with the critical exponent $\bar{\beta}=1/2$. Obtained numerical data are fit with $\Delta q\propto(U_1^c-U)^{1/2}$, where the location of the critical point is fixed to the value extracted from the fit of the correlation length for $N=1201$ (see Fig.\ref{fig:PT}(b)); so the only fitting parameter for $\Delta q$ is the non-universal pre-factor. The numerical data are in spectacular agreement with this theory prediction. Numerical data for other cuts through the Pokrovsky-Talapov transition can be found in the Appendix.

\section{Discussion}
\label{sec:conclusion}

To summarize, the nature of the quantum phase transition of the fermionic chain with nearest- and next-nearest-neighbor interactions changes from the Kosterlitz-Thouless type realized for small $U_2$ to the Pokrovsky-Talapov commensurate-incommensurate transition realized when $U_2$ is strong. Incommensurate oscillations appears in the Luttinger liquids ubiquitously by tuning the total density with the chemical potential. In the present model we witness  a different mechanism - an incommensurability appears due to a competition between the next-nearest-neighbor repulsion and the imposed constraint on the filling. This opens new possibilities in the theory of Mott-U transitions in the presence of competing interactions or frustration.

What are the consequences of the floating phase and the commensurate-incommentusrate Pokrovsky-Talapov transition? First of all, the very fact that Kosterlitz-Thouless transition turns into a Pokrovsky-Talapov transition is surprising and to the best of our knowledge such possibility has not been reported yet neither in the context of field theory nor in the framework of lattice models.  
 Furthermore, on the Luttinger liquid side of the Pokrovsky-Talapov transition the critical value of the Luttinger liquid parameter is equal to $K_\mathrm{PT}=1/p^2=1/9$, where $1/p$ is the maximal density of the corresponding commensurate phase\cite{Pokrovsky_Talapov,verresen}.
 However, at the Kosterlitz-Thouless transition as we have seen $K^c=2/9$.  The mismatch between the two critical exponents implies that equal-K lines with $1/9<K<2/9$ will condensate at the point where the nature of the transition changes as sketched in Fig.\ref{fig:phasediag}(b). Qualitatively, slightly convex curvature of the critical line for large $U_2$ agree with this picture.
   However, quantitative verification of this prediction  would require much longer chains such that in the direct vicinity of the Pokrovsky-Talapov transition the chain will host multiple helices to have a reliable fit. Because of strong frustration and low-lying excitations the convergence in the Luttinger liquid phase for large $U_2$ is extremely slow, while the accuracy required to capture the floating phase must be kept high. This question is left open for future investigations.  It would be very interesting and instructive to have insights from the field theory on the nature of the turning point at which the transition changes its nature.

It should be possible to directly program the fermionic model of Eq.(\ref{eq:ferm}) at one-third filling in optical cavities with individual control over the trapped atoms. This will allow to probe both, the Kosterlitz-Thouless and the Pokrovsky-Talapov transitions experimentally. Also note that floating phase in the vicinity of the Pokrovsky-Talapov transition, more specifically the floating phase with the Luttinger liquid parameter $1/9<K<1/8$\cite{verresen}, is stable against a single-particle instability unlike the rest of the Luttinger liquid phase on the phase diagram. This provides an alternative way to probe the change of nature of the quantum phase transitions in experiments.

\section{Acknowledgments}
I am indebted to Frederic Mila for extensive discussions on related projects. This research has been supported by Delft Technology Fellowship. Numerical simulations have been performed at the Dutch national e-infrastructure with the support of the SURF Cooperative.

\begin{appendix}

  \section{Additional data for the Pokrovsky-Talapov transition}
  
\begin{figure}[t!]
\centering 
\includegraphics[width=0.45\textwidth]{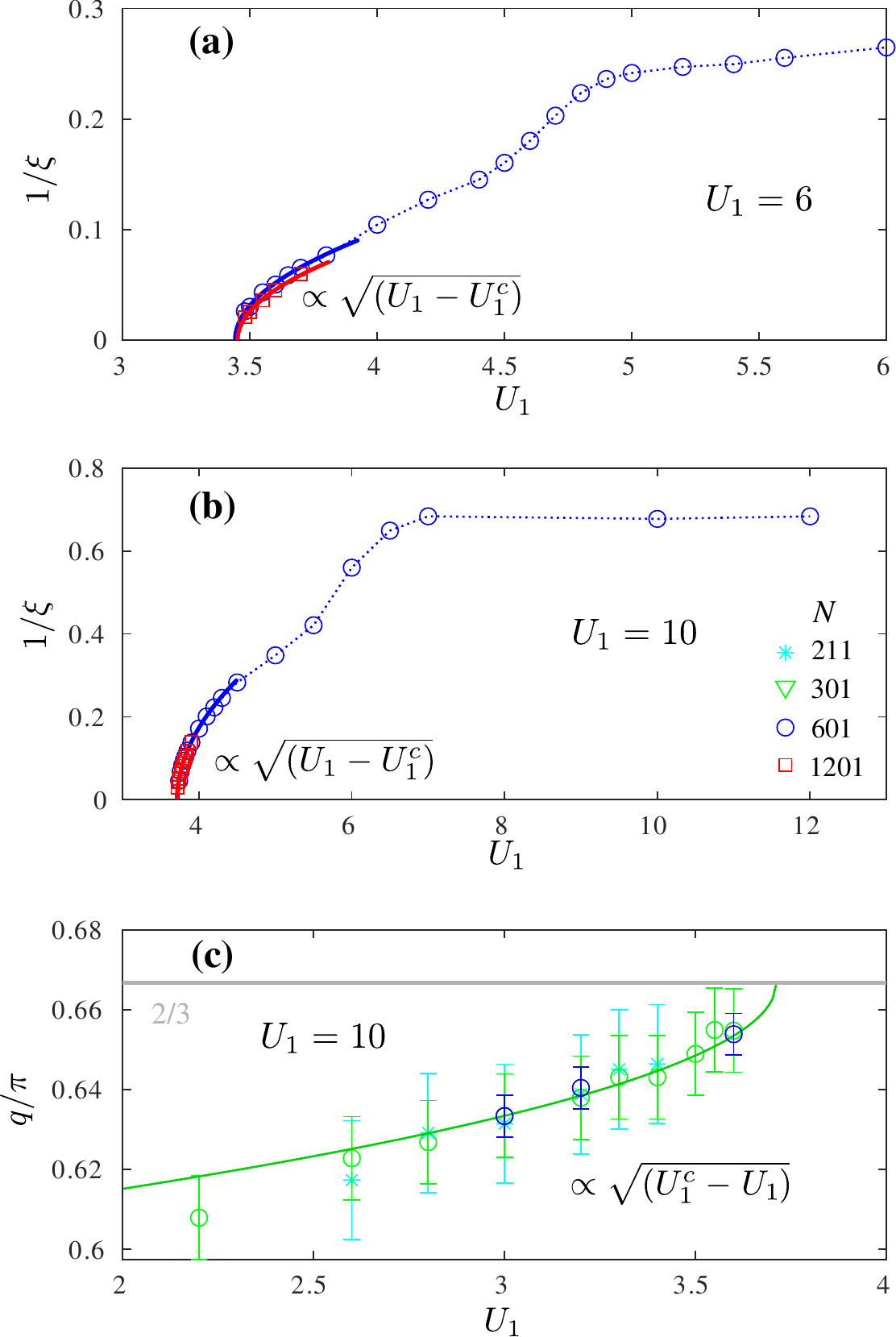}
\caption{ Additional numerical data across the Pokrovsky-Talapov transition. (a),(b) Inverse of the correlation length upon approaching the transition out of the period-3 phase at (a) $U_2=6$ fairly close to the turning point and (b) $U_2=10$. Symbols are numerical data, lines are fits with $1/\xi\propto (U_1-U_1^c)^{\nu}$ with the Pokrovsky-Talapov critical exponent $\nu=1/2$. One can see the appearance of pronounced kinks inside the period-3 phase. The location of these kinks corresponds to green squares on the phase diagram of Fig.\ref{fig:phasediag}. (c) Incommensurate wave-vector $q$ as a function of $U_1$ upon approaching the Pokrovsky-Talapov transition. Numerical data (symbols) agree with the theory prediction (line) $q-2\pi/3 \propto (U_1^c-U_1)^{\bar{\beta}}$ with the Pokrovsky-Talapov critical exponent $\bar{\beta}=1/2$. Green line show the result of the fit for $N=301$ with  $U_1^c$ being  fixed to the value obtained in (b) for $N=1201$ sites. }
\label{fig:addpt}
\end{figure}

   In this appendix we provide additional data for the Pokrovsky-Talapov transition across the cuts  at $U_2=6$ (Fig.\ref{fig:addpt}(a)) and at $U_2=10$ (Fig.\ref{fig:addpt}(b-c)). 
   Along both cuts the correlation length diverges with the Pokrovsky-Talapov critical exponent $\nu=1/2$. For $U_2=10$ one can also compare the data of the incommensurate wave-vector $q$ with theory expectation $q/\pi\propto (U_1^c-U_1)^{\bar{\beta}}$, where the location of the critical point has been extracted by sitting the correlation length as shown in Fig.\ref{fig:addpt}(b) and the critical exponent is fixed to the value of the Pokrovsky-Talapov transition $\bar{\beta}=1/2$. For both values of next-nearest-neighbor coupling $U_2$ one can see a pronounced kink in the correlation length inside the period-three phase signaling the crossover that leads to the change of the nature of the quantum phase transition. The results agree with those presented in Fig.\ref{fig:PT}.

\end{appendix}

\bibliographystyle{apsrev4-1}
\bibliography{bibliography}

\end{document}